\documentclass[aps,prb,twocolumn,floatfix,superscriptaddress,showpacs,showdate]{revtex4}

\usepackage{graphicx}
\usepackage{bbm}
\usepackage{amsmath,amssymb}
\usepackage{color,calc}

\newcommand{\be}{\begin{equation}}
\newcommand{\ee}{\end{equation}}
\newcommand{\bea}{\begin{eqnarray}}
\newcommand{\eea}{\end{eqnarray}}
\newcommand{\nn} {\nonumber}
\renewcommand{\vr} {{\bf r}}

\newcommand{\Tr}{ {\rm Tr} \, }

\def\a{\alpha}

\def\g{\gamma}
\def\G{\Gamma}
\def\d{\delta}
\def\D{\Delta}

\def\ve{\varepsilon}

\def\L{\Lambda}

\def\S{\Sigma}

\def\vf{\varphi}
\def\F{\Phi}

\def\w{\omega}

\def\bra{\langle}
\def\ket{\rangle}

\def\xc{{\rm xc}}
\def\x{{\rm x}}

\def\Tr{{\rm Tr}\,}

\begin{document}
\title{A discontinuous functional for linear response time-dependent density functional theory: 
the exact-exchange kernel and approximate forms}
\author{Maria Hellgren}
\affiliation{International School for Advanced Studies (SISSA), via Bonomea 265, 34136 Trieste, Italy}\affiliation{Max-Planck Institute of Microstructure Physics, Weinberg 2, 06120 Halle, Germany}\author{E.\!\! K.\!\! U. Gross}
\affiliation{Max-Planck Institute of Microstructure Physics, Weinberg 2, 06120 Halle, Germany}  
\date{\today}
\begin{abstract}
We present a detailed study of the exact-exchange (EXX) kernel of time-dependent density functional theory
with an emphasis on its discontinuity at integer particle numbers. It was recently found that 
this exact property leads to sharp peaks and step features in the kernel that diverge in the dissociation limit of diatomic systems [Hellgren and Gross, Phys. Rev. A, 022514 (2012)]. To further analyze the discontinuity of the kernel we here 
make use of two different approximations to the EXX kernel: the PGG approximation and a common energy denominator  approximation (CEDA). It is demonstrated that whereas the PGG approximation neglects the discontinuity the CEDA includes it explicitly. By studying model molecular systems it is shown that the so-called field counter-acting effect in the density functional description of molecular chains can be viewed in terms of the discontinuity of the static kernel.
The role of the frequency dependence is also investigated, highlighting its importance for long-range 
charge transfer excitations as well as inner-shell excitations. 
\end{abstract}
\maketitle
\section{Introduction}
Time-dependent density functional theory (TDDFT) in its linear response (LR) formulation is a formally exact and computationally efficient method for calculating excited state properties of many-electron systems.\cite{ullbook} In addition, via the adiabatic connection fluctuation dissipation (ACFD) formula, LR-TDDFT provides a promising 
approach for determining ground-state properties.\cite{,hvb08,hvb10,hg10,gd12,ot12,ebf12} 

In TDDFT the dynamical many-electron density is calculated from a 
fictitious non-interacting system in which the electrons move in an effective 
time-dependent Kohn-Sham (KS) potential.\cite{rg84} The KS potential is the sum of the external, the Hartree and the 
unknown exchange-correlation (XC) potential, $v_\xc(\vr t)$. The latter is a unique functional of the density
and contains all the many-body effects beyond the Hartree level. To linear order in an external 
perturbing field only its first variation, evaluated at the ground-state density, $n_0(\vr)$, 
is required. The central quantity to approximate in LR-TDDFT is thus 
\be
f_\xc(\vr ,\vr',t-t' )=\left.\frac{\d v_\xc(\vr t)}{\d n(\vr' t')}\right |_{n=n_0},
\ee
known as the XC kernel.\cite{gk85}

Most approximations to $f_\xc$ are derived from functionals constructed for the ground-state energy. In this way  
all history dependence is neglected and $f_\xc$ becomes independent of frequency. These so-called 
adiabatic approximations work rather well in many cases. Low-lying molecular excitation energies 
can be determined quite accurately and even true many-body features such as resonances in the optical spectrum 
due to the mixing of discrete single-particle states with continuum states are qualitatively described.\cite{sdl95,sdg01,hvb09} Different studies also show that ground-state properties within the ACFD framework are relatively insensitive to the lack of frequency dependence.\cite{lgp00,hvb10} On the other hand, in the important and challenging cases of charge-transfer, inner-shell and double excitations a proper frequency dependence must be included.\cite{mzcb04,mt06,hg12} 

Many works have focused on understanding what features $v_\xc$ and $f_\xc$ must have in order to capture different many-body effects. Such an important exact property of $v_\xc$, discovered already in the ground-state theory, is the so-called derivative discontinuity.\cite{pplb82} The ground-state XC energy as a function of particle number exhibits kinks at the integers which reflects the fact that the energy levels of the KS system do not correspond to the true energy levels. In particular, the KS affinity, i.e. the lowest unoccupied KS level, must be corrected with an amount exactly equal to the size of the derivative discontinuity in order to get the true affinity.

A kink in the XC energy leads to a discontinuity in $v_\xc$ in terms of a constant shift.  Although an overall shift cannot affect the density there are fundamental consequences of the discontinuity also for the density. It can appear in subregions of space where the density integrates to a number close to an integer. There it forms steps which act to prevent delocalization of the charges. This important effect is most clearly demonstrated when breaking chemical bonds.\cite{perde,mkk11}
 
For time-dependent problems very similar steps in the XC potential become important in many situations.\cite{efmr13} For example, it has been shown to be a crucial feature for describing the Coulomb blockade effect in the time-domain\cite{coublock} as well as for accurately reproducing ionization processes.\cite{lk05} 

The linear density response function, $\chi$, is within TDDFT determined from the KS non-interacting response function $\chi_s$ according to a Dyson-like equation\cite{pgg96}
\be
\chi(\w)=\chi_s(\w)+\chi_s(\w)[v+f_\xc(\w)]\chi(\w).
\ee
where all quantities are matrices in $\vr,\vr'$ and $v$ is the bare Coulomb interaction.
In insulating solids, a difficulty when calculating the optical spectra from $\chi$ is to incorporate the correct band gap, since the KS gap, contained in $\chi_s$, in general is much smaller than the true gap.\cite{gruning} Since the gap is given by the difference between the ionization energy and the affinity, it is expected that the information on how to correct the gap lies in the derivative discontinuity. However, in LR-TDDFT that information has to be carried by the XC kernel. 
A similar situation occurs in the case of charge-transfer (CT) excitations.\cite{tozer} If one electron is transferred between two 
fragments, in the limit of infinite separation, the excitation energy is given by the difference in the ionization energy of the donor and the affinity of the acceptor. Again, the latter has to be corrected with the derivative discontinuity. In the case of CT excitations it has been shown that this correction comes either entirely from the XC kernel or from a combination of the XC potential and the XC kernel.\cite{hg12}

The question then naturally arises, how does the derivative discontinuity affect the XC kernel? 
In a recent paper\cite{hg12} the present authors investigated this question and found a discontinuity of XC kernel with spatial divergencies. Moreover, it was found that the discontinuity could have a strong frequency dependence. The purpose of the present paper is to further analyze as well as to present some more examples where the discontinuity of the XC kernel plays an important role. 

A framework used for constructing functionals with the derivative discontinuity and with frequency dependence 
is the variational formulation of many-body perturbation theory (MBPT).\cite{vbdvls05,casida95} 
An advantage of the MBPT approach to TDDFT is that the relevant physics can be built into the functional 
via the intuitive Feynman diagram expansions. The first approximation is the so-called time-dependent 
exact-exchange (TDEXX) approximation, which in MBPT corresponds to the Klein functional at the level of the
time-dependent Hartree-Fock (TDHF) approximation. The TDEXX has already been used to calculate many 
different properties of atoms, molecules and solids.\cite{kg02,hvb09,hvb10,hg10,hg11m,hg11} With regards to spectral 
properties it has some limitations due to the double inversion of the KS density response function \cite{hvb09} 
but for ground-state properties within the ACFD framework TDEXX has produced excellent results in 
terms of total energies, polarizabilities and van der Waals coefficients.\cite{hvb08,hvb10,hg10,hg11} 
Furthermore, the static EXX potential exhibits a discontinuity which carries over to the time dependent potential as shown in Ref. \onlinecite{mk05}. 

In the present paper we will study the discontinuity of the TDEXX kernel in order to demonstrate some of its fundamental properties. Apart from studying the exact TDEXX kernel 
we also employ two different approximations: a Slater-type of approximation which is also known as the PGG 
approximation,\cite{pgg96} and a KLI-type\cite{kli92} of approximation previously derived in Refs. \onlinecite{gb01,casida96}. The latter is here implemented for the first time allowing for numerical comparisons.
It will be shown that the PGG kernel completely neglects the discontinuity whereas the KLI approximation incorporates it explicitly. 

The paper is organized as follows. We start with a review of the discontinuities in ensemble DFT and then we 
generalize the discussion to the time-dependent case and derive the discontinuity of the XC kernel. 
In Sec. III we present the MBPT framework. In Sec. IV we derive 
the Slater and the KLI approximation to the TDEXX kernel. Then finally, in Sec. V, we present numerical results for different 1D soft-coulomb systems.  
\section{Fractional charges in DFT}
For functional derivatives to be uniquely defined and for treating densities that integrate to a non-integer 
number of electrons ensembles must be introduced.\cite{pplb82,pl97} We will start this section by reviewing 
ensembles for ground-state density functionals and derive simple formulas for evaluating the discontinuities of a 
given functional. Then, we will generalize the ensembles to treat systems in time-varying fields and 
determine the discontinuities of the dynamical XC kernel.
\subsection{Ground-state ensembles}
For an average number of electrons $N=N_0+p$, where $N_0$ is an integer it is sufficient to include two members in the ensemble\cite{pplb82} 
\be
\hat{\g}^{>}= (1-p)|\Psi_{N_0}\ket\bra\Psi_{N_0}|+p|\Psi_{N_0+1}\ket\bra\Psi_{N_0+1}|,
\label{glarge}
\ee 
where $|\Psi_{k}\ket$ is the ground-state with $k$ particles. Similarly, for $N=N_0-1+p$ we can define
\be
\hat{\g}^{<}= (1-p)|\Psi_{N_0-1}\ket\bra\Psi_{N_0-1}|+p|\Psi_{N_0}\ket\bra\Psi_{N_0}|.
\label{gsmall}
\ee 
The ensemble ground-state energy consists of straight line segments 
\bea
E^{>}&=&(1-p)E_{N_0}+pE_{N_0+1}\\
&&\nn\\
E^{<}&=&(1-p)E_{N_0-1}+pE_{N_0}
\eea
with a derivative discontinuity at the integer $N_0$ given by $I-A$, where the ionization energy is $I=E_{N_0-1}-E_{N_0}$ and the affinity is $A=E_{N_0}-E_{N_0+1}$.

In this context the KS system is defined to be the fictitious system of non-interacting electrons that can produce the same ensemble density. The KS ensemble density is given by
\bea
n^{>}(\vr)&=&\sum_k^{N_0}|\vf^{p}_k(\vr)|^2+p|\vf^{p}_{N_0+1}(\vr)|^2\\
n^{<}(\vr)&=&\sum_k^{N_0-1}|\vf^{p}_k(\vr)|^2+p|\vf^{p}_{N_0}(\vr)|^2
\eea
where the superscript $p$ is attached to denote the fact that the KS potential that determines the 
orbitals will depend on the average number of particles. The total ground state ensemble energy can be written as
\bea
E[n]&=&T_s[n]+\frac{1}{2}\int d\vr d\vr'  n(\vr)v(\vr,\vr')n(\vr')\nn\\
&&+\int d\vr\, w(\vr) n(\vr)+E_\xc[n]
\eea
where $T_s$ is the non-interacting kinetic energy functional and $E_\xc$ is the XC energy. In order to exhibit the derivative discontinuity of $E_\xc$ we take the derivative with respect to the number of particles
\bea 
\frac{\partial E}{\partial N}=\frac{\partial T_s}{\partial N}+\int d\vr\left [w(\vr)+v_{\rm H}(\vr)\right ] f(\vr)+\frac{\partial E_\xc}{\partial N},
\eea
where we have identified the Fukui function
\be
f(\vr)=\frac{\partial n(\vr)}{\partial N}.
\ee
The derivative of the kinetic energy can easily be evaluated once written in 
terms of occupied KS eigenvalues $\ve_k$. Let us first focus on an ensemble of the form of Eq. (\ref{glarge}), i.e., with 
$N=N_0+p$. The derivative with respect to $N$ is equal to the derivative with respect to $p$ and we find
\bea
\frac{\partial E^>}{\partial N}=\ve_{N_0+1}-\int d\vr\, v_\xc(\vr) f(\vr)+\frac{\partial E_\xc}{\partial N}=\ve_{N_0+1},
\label{chemeig1}
\eea
where we have used the identity
\bea
\frac{\partial E_\xc}{\partial N}=\int d\vr \,v_\xc(\vr) f(\vr)
\label{parexc}
\eea
and the definition of the XC potential $v_\xc=\d E_\xc/\d n$. The same steps can be performed for the ensemble in Eq. (\ref{gsmall}) and we find similarly 
\bea
\frac{\partial E^<}{\partial N}=\ve_{N_0}.
\label{chemeig2}
\eea
Eqs. (\ref{chemeig1}) and (\ref{chemeig2}) thus prove that the highest occupied eigenvalue must be equal to the chemical potential\cite{pl97} and should not change with $N$. Many problems with existing functionals are related to a lack of this straight-line behavior when extended to fractional charges.\cite{Cohensci,cmsy09,gould13,kk13} In the limit $N\to N_0^\pm$ we have
\bea
\left.\frac{\partial E}{\partial N}\right|_+=\ve_{N_0+1}^+,\,\,\,\, \,\,\,\left.\frac{\partial E}{\partial N}\right|_-=\ve_{N_0}^-.
\label{deriveplus}
\eea
In this limit $\ve_{N_0+1}^+=\ve^+_{\rm LUMO}$, i.e., the lowest unoccupied KS orbital obtained from the KS potential in the limit $N\to N_0^+$ ($V_s^+$). In the same way $\ve_{N_0}^-=\ve^-_{\rm HOMO}$, i.e., the highest occupied KS orbital obtained from $V_s^-$. It is important from which direction the limit is taken since $v_\xc$ has a discontinuity at $N=N_0$. The discontinuity in $v_\xc$ is in general positive, shifting the KS affinity $A_s=-\ve^-_{\rm LUMO}$ to the true affinity $A=-\ve^+_{\rm LUMO}$ in order to obey the relation in Eq. (\ref{chemeig1}). 

The discontinuity of $v_\xc$ is related to the derivative discontinuity in $E_\xc[n[w,N]]$. We can use the identity in Eq. (\ref{parexc}) to formally express the value of the discontinuous shift $\D_\xc$ at $N_0$. For $N>N_0$ we write $v^>_{\xc}(\vr)=v^-_{\xc}(\vr)+\D_\xc(\vr)$ and insert into Eq. (\ref{parexc}) 
\bea
\frac{\partial E^>_{\xc}}{\partial N}=\int d \vr\,\,  \left[v^-_{\xc}(\vr)+\D_\xc(\vr)\right]f(\vr).
\label{forstaderivN}
\eea
Taking the limit $N\to N_0^+$ and by rearranging we find
\bea
\D_\xc=\left.\frac{\partial E_{\xc}}{\partial N}\right |_+-\int d \vr\, v^-_{\xc}(\vr)f^+(\vr),
\label{forstaderivN2}
\eea
where we have used the fact that the Fukui function integrates to unity. If $\partial E_{\xc}/\partial N$ has a non-trivial discontinuity at $N_0$, $\D_\xc$ is finite.
A discontinuous shift in the XC potential also implies a shift in the eigenvalues with the same magnitude. We can thus write Eq. (\ref{deriveplus}) in the previous section as
\bea
-A=\left.\frac{\partial E}{\partial N}\right|_+=\ve_{\rm LUMO}^-+\D_\xc.
\eea
 
In order to determine the discontinuities of the XC kernel we start by noting that for particle number conserving variations of the density $f_\xc$ is only defined up to the sum of two arbitrary functions $g_\xc(\vr)+g_\xc(\vr')$. This observation follows immediately after inspecting the definition of $f_\xc$
\bea
\d^2E_\xc=\int d\vr'd\vr f_\xc(\vr,\vr')\d n(\vr')\d n(\vr).
\eea
When we instead allow for arbitrary density variations $f_\xc$ becomes unique but may have discontinuities of the form 
\bea
f^+_\xc(\vr',\vr)-f^-_\xc(\vr',\vr)=g_\xc(\vr)+g_\xc(\vr').
\eea
In order to evaluate $g_\xc$ given a functional $E_\xc$ we use the same procedure as for the XC potential. Let us study the quantity
\bea
\frac{\d}{\d w(\vr)}\frac{\partial E_\xc}{\partial N}=\int\! d \vr d \vr'\,\chi(\vr_1,\vr) f_\xc(\vr,\vr')f(\vr')\nn\\
+\!\int \!d \vr \,v_\xc(\vr)\frac{\d f(\vr)}{\d w(\vr_1)}.
\label{fxder}
\eea
Writing $f^>_\xc(\vr',\vr)=f^-_\xc(\vr',\vr)+g_\xc(\vr',\vr)$ and taking the limit $N\to N^+_0$, $g_\xc(\vr',\vr)\to g_\xc(\vr)+g_\xc(\vr')$ and we arrive at
\begin{widetext}
\bea
\int d \vr \,\chi(\vr_1,\vr) g_\xc(\vr)=\left.\frac{\d}{\d w(\vr)}\frac{\partial E_\xc}{\partial N}\right|_+
-\!\int\! d \vr d \vr'\,\chi(\vr_1,\vr) f^-_{\xc}(\vr,\vr')f^+(\vr')-\!\int \!d \vr \,
v^+_\xc(\vr)\frac{\d f^+(\vr)}{\d w(\vr_1)}.
\label{muxcdeltan0}
\eea
\end{widetext}
This equation only determines $g_\xc$ up to constant. The constant can, however, be fixed by considering the the second derivative of $E_\xc$ with respect to $N$
\bea
 2\int d \vr \,f^+(\vr)g_\xc(\vr)&=&\left.\frac{\partial^2 E_\xc}{\partial N^2}\right|_+-\int  d \vr' \,v^+_\xc(\vr')\frac{\partial f^+(\vr')}{\partial N}
 \nonumber \\
&&\!\!\!\!\!\!\!\!\!\!\! -\int d \vr d\vr' \, f^+(\vr)f^-_{\xc}(\vr,\vr')f^+(\vr'),
\label{muxcdeltandeltan}
\eea
yielding a condition to be imposed on Eq. (\ref{muxcdeltan0}). The function $g_\xc$ obtained via Eqs. (\ref{muxcdeltan0}-\ref{muxcdeltandeltan}) 
was recently analyzed in Ref. \onlinecite{hg11} showing a diverging behavior of the form 
\be
g_\xc(\vr)\sim \frac{|\vf_{N_0+1}(\vr)|^2}{n(\vr)}\sim e^{2(\sqrt{2I}-\sqrt{2A_s})\, r},
\ee
as $r\to \infty$. In the next section we will generalize the these ideas to TDDFT.
\subsection{Time-dependent ensembles}
The discontinuity, found as an exact property in the ground-state XC potential, appears also in the time-dependent XC potential. In, e.g., an ionization process the particle number will change locally on the molecule and hence $v_\xc$ around the molecule will be 
evaluated close to an integer.\cite{lk05} In the case of quantum transport, electrons are transferred from a lead to a weekly connected central region, 
which could be a quantum dot or a molecule. In order to describe the Coulomb blockade effect, it has been shown that in the central region $v_\xc$ forms a step, which has been associated with the derivative discontinuity.\cite{coublock} 

In the time-dependent case the discontinuity will be rather different from the ground-state discontinuity. The size of the jump will depend on the density at the time when the particle number crosses an integer. That density may not be the ground-state density. Furthermore, a history dependence could be important.\cite{efmr13} As a consequence, in TDDFT 
the analysis of the discontinuities of a given functional becomes much more complicated. In this work we will, however, 
not aim for such a general description of the discontinuity of the time-dependent XC potential. Instead we will focus only on the linear response regime, in which the kernel depends only on time differences. Numerical results suggest that the discontinuity of the XC kernel carries a frequency-dependence\cite{hg12,efmr13}, which indicates that memory effects 
make a difference. To see that a frequency dependence in principle is allowed for, we examine the definition of the 
dynamical kernel for particle number conserving density variations $\d n(\vr,\w)$ 
\be
\d v_\xc(\vr, \w)=\int d\vr' f_\xc(\vr, \vr', \w)\d n(\vr' \w). 
\ee
From this definition, we see that we can always add two functions $g_\xc(\vr, \w)$ and $g_\xc(\vr', \w)$ to $f_\xc$ without 
changing the physical results obtained from the density response function. The function $g_\xc(\vr, \w)$ will clearly vanish when 
integrating over $\vr'$ and the function $g_\xc(\vr', \w)$ will merely generate an irrelevant constant to $v_\xc$.
In order to make the functional derivative unique we thus need to allow for variations that can change the particle numbers. 

Given a general functional $F[n]$ its functional derivative with respect to $n(\vr t)$ is defined as
\be
\d F[n]=\int \!d\vr dt\, \frac{\d F}{\d n(\vr t)}\d n(\vr t),
\ee
where the integral over time is taken over the interval $[0,T]$. If the density variations conserve the particle numbers, $\d F/\d n(\vr t)$ is only defined up to a function $C(t)$. Allowing a change of particle numbers that is constant in time remove some of the arbitrariness but leaves the derivative still undefined up to a function $S(t)$
with the property
\be
\int \!dt\, S(t)=0.
\ee
We thus see that we have to allow the particle number to change in time in order to completely fix the 
functional derivatives, which suggests that we need ensembles that varies the particle number in time in order 
to exhibit the discontinuities of a given functional. To see this more clearly we start by defining ensembles as in Eqs. (\ref{glarge}-\ref{gsmall}) but now replacing the ground-states of the $N$ and $N+1$ particle systems with states that evolve in time in the external potential $w(t)$. Let us now consider the $N$-derivative of the XC potential, defined on the ensemble densities, and evaluate it at the ground-state density with $N=N_0^+$, i.e., $n(\vr t)=n_0^+(\vr)$ 
\bea
\left.\frac{\d v_\xc(\vr t)}{\d N}\right |_{n^+_0}=\int d\vr'dt' f^-_\xc(\vr,\vr', t-t')f^+(\vr')\nonumber\\
+\int d\vr' dt' g_\xc(\vr', t-t')f^+(\vr')+\int dt' g_\xc(\vr, t-t').
\eea
Clearly, this equation does not allow us to completely determine the discontinuity $g_\xc$. 
We therefore propose an ensemble where also the coefficients vary in time and thus allows the particle 
number to change as a function of time. We write
\bea
\hat{\g}^{>}(t)&=&\{1-p_1(t)\}|\Psi_{N_0}(t)\ket\bra\Psi_{N_0}(t)|\nn\\
&&\,\,\,\,\qquad+p_1(t)|\Psi_{N_0+1}(t)\ket\bra\Psi_{N_0+1}(t)|
\label{ens1}
\eea 
\bea
\hat{\g}^{<}(t)&=&\{1-p_2(t)\}|\Psi_{N_0-1}(t)\ket\bra\Psi_{N_0-1}(t)|\nn\\
&&\,\,\,\,\qquad+p_2(t)|\Psi_{N_0}(t)\ket\bra\Psi_{N_0}(t)|,
\label{ens2}
\eea 
where $p_{1}(t)$ and $p_{2}(t)$ are two arbitrary functions that can vary between 0 and 1. The time-dependent number of 
particles are then $N^{>}(t)=N_0+p_1(t)$ and $N^{<}(t)=N_0-1+p_2(t)$, respectively. We are only interested in linear response which 
means that the functionals should be evaluated at the ground-state density. Assuming that $p_1(t_0)=0^+$ i.e. $N(t_0)=N_0^+$, 
where $t_0$ is the initial time, we compare this kernel $f^+_\xc(\vr,\vr',\w)$ to the one evaluated at $p_2(t_0)=1^-$ i.e. 
$N(t_0)=N_0^-$ denoted $f^-_\xc(\vr,\vr',\w)$. These kernels are thus evaluated at the same ground-state density, 
the only difference being from which side of the integer the limit is taken. Now let us use Eq. (\ref{ens1}), take the derivative of $v_\xc$ with respect to the time dependent number of particles and evaluate it at $n_0^{+}=n^+(\vr t_0)$
\bea
\left.\frac{\d v_\xc(\vr t)}{\d N(t')}\right |_{n^+_0}=\int d\vr' f^-_\xc(\vr,\vr', t-t')f^+(\vr')\nonumber\\
+\int d\vr' g_\xc(\vr', t-t')f^+(\vr')+g_\xc(\vr, t-t').
\label{fircond}
\eea
The function $g_\xc(\vr,\w)$ can now be determined. The ensemble proposed thus allow functional derivatives 
to be uniquely defined. 

In the next section we will study $g_\xc(\vr,\w)$ within the TDEXX approximation. 
\section{Restricted Klein functional}
In this section we will introduce a framework for constructing advanced 
functionals for DFT and TDDFT.\cite{vbdvls05} The basic idea is to use the Klein action 
functional\cite{klein1961} formulated in terms of the many-body Green's function $G$, 
and then to restrict the variational freedom to KS Green's functions
$G_s$ coming from a local multiplicative KS potential $V$. 

The Klein functional is defined as 
\be
 i Y_{\rm K}[G]=\F[G]-\Tr\left\{
GG_{\rm 0}^{-1}-1+\ln(-G^{-1})\right\},
\label{kleinf}
\ee
where $G_0$ is the non-interacting 'bare' Green's function carrying information about the external potential $w$ and the number of particles. The trace $\Tr$ denotes a sum over one electron states plus an integral over the Keldysh contour.\cite{ldsab05} 
The functional $\F$ is constructed such that the self-energy $\S$ is given by
\be
\Sigma=\frac{\delta\F}{\delta G}.
\ee
Varying the Klein functional with respect to $G$ yields the Dyson equation 
\be
G=G_{\rm 0}+G_{\rm 0}\S G,
\label{dys}
\ee
as a condition that renders the functional stationary. Furthermore, for static problems the stationary point of the Klein functional is equal to the total energy $iY_{\rm K}=E^{\rm tot}$. Not all possible approximate self-energies can be constructed from a $\F$ functional as can easily be verified to third order in the Coulomb interaction. The set of those self-energies that can are called 
$\Phi$-derivable or {\em conserving} approximations  since it can be shown that the resulting $G$ incorporates basic conservation 
laws like energy, particle and momentum conservation.\cite{baym} 

By restricting the variational freedom to KS Green's functions we simply replace $G$ by $G_s$ in Eq. (\ref{kleinf}). 
The non-interacting $G_s$ is easily constructed and in the static case the Klein functional 
simplifies to 
\be
Y_{\rm K}[V]= T_s[n] + \int \!\!d\vr\, w(\vr) n(\vr)-i \F[G_{s}] ,
\label{energi}
\ee
where $T_s$ is the kinetic energy of non-interacting electrons in the potential $V$. The $\F$-functional thus 
plays the role of the Hartree and XC energy.\cite{casida95,vbdvls05} In this way we can generate approximate functionals in DFT and TDDFT from $\F$-derivable self-energies in MBPT. These functionals are implicit, nonlocal density functionals via $G_s$, which depends on KS orbitals and eigenvalues. Therefore, they can overcome some of the limitations involved in local and explicit functionals of the density. 
Moreover, the hope is that the physics included in the self-energy is reflected in the performance of the density functional as well. The level to which this is true has recently been investigated.\cite{fhiprl} 

We can now use standard TDDFT results to determine the XC potential $v_{\xc}$. 
Using the chain rule $\d \F^\xc/\d G_s*\d G_s/\d V=\d \F^\xc/\d n*\d n/\d V$ we find 
the so called linearized Sham-Schl\"uter (LSS) equation\cite{ShamSchluter} (with $\vr_1 t_1\rightarrow 1$):
\be
\int \chi_s(1,2)v_{\xc}(2)d2=\int \S^\xc_s(2,3)\Lambda(3,2;1)d2d3,
\label{lsseq}
\ee
where $\S_s^\xc$ is the self-energy calculated with KS orbitals generated by $v_\xc$ and 
$$
i\Lambda(3,2;1)=\frac{\delta G_s(3,2)}{\delta V(1)}=G_s(3,1)G_s(1,2).
$$
A further variation of the LSS equation with respect to the potential $V$ 
results in an equation for $f_\xc$:
\begin{eqnarray}
&&\int \chi_s(1,2)f_{\xc}(2,3)\chi_s(3,4)d2d3\nn\\
&&\,\,\,\,\,\,\,\,\,\,=\int \frac{\delta\S^\xc_s(2,3)}{\delta 
V(4)}\Lambda(3,2;1)d2d3\nn\\
&&\,\,\,\,\,\,\,\,\,\,\,\,\,\,+\int 
\Lambda(1,2;4)D(2,3)G_s(3,1)d2d3\nn\\
&&\,\,\,\,\,\,\,\,\,\,\,\,\,\,+\int 
G_s(1,2)D(2,3)\Lambda(3,1;4)d2d3,
\label{fxceq}
\end{eqnarray}
where $D(2,3)=\S^\xc_s(2,3)-v_{{\xc}}(2)\delta(2,3)$. 

The XC functional derived from the Klein functional approach has been shown to generate XC potentials which has many features of the exact XC potential. Of particular interest for this work is the derivative discontinuity, which is well reproduced at least at even integer electron number. 

In order to investigate these functionals for densities that integrate to a non-integer particle number we will here assume that the proper generalization is made by inserting an ensemble KS Green's function $G_s^E$ constructed from the ensembles introduced in Sec. II (see Ref. \onlinecite{hg12}). This has proven to be the right procedure using HF and GW self-energies.\cite{hrg12} Using Eq. (\ref{forstaderivN2}) we can express the discontinuity of the ensemble $v^E_\xc$ as\cite{pplb82,hg12}
\bea
\D_\xc&=&\int d\vr d\vr' \vf_{\rm L}(\vr)\left[\S^\xc_s[G_s^+](\vr,\vr',\ve^+_{\rm L})\right.\nn\\
&&\,\,\,\,\,\,\left.-v^-_\xc(\vr')\d(\vr-\vr')\right]\vf_{\rm L}(\vr'),
\label{perdew}
\eea
where $\vf_{\rm L}$ is the LUMO orbital of the $N^-$-system and $\ve^+_{\rm L}=\ve^-_{\rm L}+\D_\xc$. At any given $N$ the ensemble $v^E_\xc$ is determined by Eq. (\ref{lsseq}) by replacing $\S^\xc_s[G_s]$ with $\S^\xc_s[G_s^E]$, $\L$ with $\L^E$ and $\chi_s$ with $\chi_s^E$.
$v^E_\xc$ is, however, not uniquely determined by Eq. (\ref{lsseq}), but only up to a constant. This constant can be fixed using Eq. (\ref{parexc}). For instance, when $N< N_0$ we find
\bea
0=\bra \vf_{\rm H}|\S^\xc_s[G_s^E](\ve_{\rm H})-v^E_\xc|\vf_{\rm H}\ket
\label{perdew2}
\eea
(in bracket notation). This condition results in $v^E_\xc(\vr)\to 0$ when $\vr\to\infty$. 

Similarly, the {\em static} ensemble XC kernel can be evaluated from the second derivative of Eq. (\ref{lsseq}), leading to an equation similar to Eq. (\ref{fxceq}) at $\w=0$.  
The non-uniqueness now amounts to a function depending on $\vr$ in both inversions of $\chi^E_s$. 
In order to fix these functions we proceed as follows.
For the first inversion Eq. (\ref{fxder}) can be used directly as a condition to impose on Eq. (\ref{fxceq}). In the second inversion we need a condition of the kind $\int d\vr f_\xc(\vr',\vr)q(\vr)=Q(\vr')$. Such condition can be found from Eq. (\ref{fxder}) by imposing Eq. (\ref{muxcdeltandeltan}) and solve for $\int d\vr f_\xc(\vr',\vr)f(\vr)$. 

In order to write an equation for the discontinuity of $f^E_\xc$ at any given frequency the time-dependent ensembles Eq. (\ref{ens1})-(\ref{ens2}) must be used. The derivative in Eq. (\ref{fircond}) can be evaluated by taking the derivative of Eq. (\ref{lsseq}), assuming $\chi_s$, $G_s$, $\L$ and $n$ are obtained as expectation values of Eq. (\ref{ens1})-(\ref{ens2}). That this generalization makes sense is not obvious and each approximation should be carefully investigated independently. However, if we apply this assumption we find the following equation determining the frequency-dependent discontinuity
\begin{widetext}
\bea
\int d \vr_2 \,\chi_s(\vr_1,\vr_2,\w) g_\xc(\vr_2,\w)&=&\left.\int \!d2 d3\,\S_s^{\rm xc}(2,3)\frac{\d \L(2,3;1)}{\d N(t)}\right|_{n^+_0}+\left.\int d(2345)\frac{\d \S_s^{\rm xc}(2,3)}{\d G_s(4,5)}\frac{\d G_s(4,5)}{\d N(t)}\L(3,2;1)\right|_{n^+_0}\nn\\
&&-\!\int\! d \vr_2 d \vr_3\,\chi_s(\vr_1,\vr_2,\w) f^-_{\xc}(\vr_2,\vr_3,\w)f^+(\vr_3)-\!\left.\int \!d 2 \,
v^+_\xc(2)\frac{\d\chi_s(1,2)}{\d N(t)}\right|_{n^+_0}.
\label{disckleintd}
\eea
\end{widetext}

In the limit $\w=0$ this equation reduces to the equation obtained by considering a static ensemble.
We note also that, in linear response, all quantities are evaluated at the ground-state and hence the precise
form of $N(t)$ in time will not enter the equation. Eq. (\ref{disckleintd}) determines $g_\xc$ up to a constant. At $\w=0$ we can use Eq. (\ref{muxcdeltandeltan}) to fix this constant. For $\w\neq 0$ it is not clear how to fix the constant, since we started from an equation that cannot completely determine the ensemble $v_\xc$. 

Until now, functionals based on the Klein expression has been studied in the HF approximation, the second order Born and the GW, corresponding to the EXX, MP2 and RPA respectively in the restricted density functional framework. 
In the time-dependent linear response regime only the TDEXX kernel has been implemented for atoms. The frequency dependence found had some deficiencies, indicating that exchange effects might have to be treated more carefully when translated from MBPT to TDDFT.
In the next section we will mainly focus on the static kernel and investigate its discontinuities in detail. This will be done with the help of common energy denominator approximations. We will also investigate the role of the frequency dependence.
\section{Exact exchange and the KLI approximation}
\begin{figure}[t]
\includegraphics[width=8.5cm, clip=true]{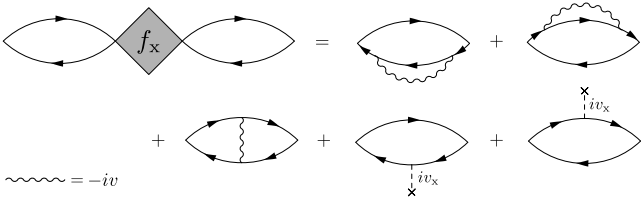}\\
\caption{Diagrammtic representation of Eq. (\ref{fxceq}) in the EXX approximation. Solid arrowed lines represent time-ordered Green's functions.}
\label{ediag1}
\end{figure}

In this section we will focus on the EXX approximation which corresponds to the HF approximation within MBPT. 
In the EXX we have
\be
\Phi^\x=\frac{i}{2}\Tr\left[G_sG_sv\right],\quad \S_s^\x(1,2)=i G_s(1,2)v(1,2).
\label{phix}
\ee 
For calculating the TDEXX kernel from Eq. (\ref{fxceq}) we also need the variation of the self-energy with respect to $V$
$$
\frac{\delta \S^{\rm x}_s(2,3)}{\delta V(4)}=-v(2,3)\Lambda(2,3;4).
$$
The terms on the right hand side of Eq. (\ref{fxceq}) can be represented diagrammatically as in Fig.~\ref{ediag1}.
Evaluating the trace in Eq. (\ref{phix}) we find the standard expression for the ground-state EXX energy 
\be
E_\x=-i\Phi_s^\x=-\int\! d\vr d\vr' \g(\vr,\vr')v(\vr,\vr')\g(\vr,\vr'),
\label{exxen}
\ee
in terms of the KS density matrix $\g(\vr,\vr')=\sum_{k}n_k\vf_k(\vr)\vf_k(\vr')$, where $n_k$ is the occupation number of orbital $k$ which can be either zero (unoccupied) or one (occupied). 
The discontinuity of the static EXX potential is equal to
\bea
\D_\x&=&\int d\vr d\vr' \vf_{\rm L}(\vr)\left[\S_\x(\vr,\vr')\right.\nn\\
&&\,\,\,\,\,\,\left.-v^-_\x(\vr')\d(\vr,\vr')\right]\vf_{\rm L}(\vr'),
\label{perdew}
\eea
where $\S_\x(\vr,\vr')=-\g(\vr,\vr')v(\vr,\vr')$.
The discontinuity of the TDEXX kernel can be determined using Eq. (\ref{disckleintd}). We find
\begin{widetext}
\bea
\int d \vr_2 \,\chi_s(\vr_1,\vr_2,\w) g_\xc(\vr_2,\w)&=&4\sum_{k\ne \rm L}\int d\vr_2d\vr_3[v(\vr_2,\vr_3)\g(\vr_2,\vr_3)-v^+_\x(\vr_2)\d(\vr_2,\vr_3)]\frac{\vf_{\rm L}(\vr_2)\vf_k(\vr_3)\vf_{\rm L}(\vr_1)\vf_k(\vr_1)\ve_{k{\rm L}}}{\w^2-\ve_{k{\rm L}}^2}\nn\\
&&\!\!\!\!\!\!\!\!\!\!\!\!\!\!\!\!\!\!\!\!\!\!\!\!\!\!\!\!\!\!\!\!\!\!\!\!\!\!\!\!\!+\frac{1}{2}\int d\vr_2d\vr_3v(\vr_3,\vr_2)\vf_{\rm L}(\vr_2)\vf_{\rm L}(\vr_3)\L(\vr_3,\vr_2;\vr_1,\w)-\!\int\! d \vr_2 d \vr_3\,\chi_s(\vr_1,\vr_2,\w) f^-_{\xc}(\vr_2,\vr_3,\w)|\vf_{\rm L}(\vr_3)|^2.
\label{disckleintdexx}
\eea
\end{widetext}
We note that this equation is the limit $N\to N_0^+$ of the equation used for uniquely determining $f_\x$.  

A very useful and accurate approximation to the EXX potential is the so-called KLI approximation.\cite{kli92} Apart from leading to a potential and kernel which may have numerical advantages the KLI approximation exhibits the discontinuity in an instructive way. The KLI approximation avoids the inversion of the full $\chi_s$ by means of a common energy denominator approximation (CEDA). The CEDA sets all KS excitation energies to the same value $\D\ve$. In the retarded $\chi_s$ this procedure amounts to the following: 
\bea
\chi_s(\vr_1,\vr_2;\omega) &=&4\sum_{kk'}n_k(1-n_{k'})\frac{\ve_{k'}-\ve_k}{\omega^2-(\ve_{k'}-\ve_k)^2}\nn\\
&&\,\,\,\,\,\,\,\,\,\,\,\,\,\,\,\times\,\vf_k(\vr_1)\vf_{k'}(\vr_1)\vf_k(\vr_2)\vf_{k'}(\vr_2)\nn\\
&\!\!\!\!\!\!\!\!\!\!\!\!\!\!\!\!\!\!\!\!\!\!\!\!\!\!\!\!\!\!\!\!\!\!\!\!\!\!\!\!\!\!\!\!\!\!\!\!\!\!\!\!\!\!\!\!\!\!\!\!\approx&\!\!\!\!\!\!\!\!\!\!\!\!\!\!\!\!\!\!\!\!\!\!\!\!\!\!\!\!\!\!\!\!A(\w)\sum_{kk'}n_k(1-n_{k'})\vf_k(\vr_1)\vf_{k'}(\vr_1)\vf_k(\vr_2)\vf_{k'}(\vr_2)\nn\\
&\!\!\!\!\!\!\!\!\!\!\!\!\!\!\!\!\!\!\!\!\!\!\!\!\!\!\!\!\!\!\!\!\!\!\!\!\!\!\!\!\!\!\!\!\!\!\!\!\!\!\!\!\!\!\!\!\!\!\!\!=&\!\!\!\!\!\!\!\!\!\!\!\!\!\!\!\!\!\!\!\!\!\!\!\!\!\!\!\!\!\!\!\!A(\w)\g(\vr_1,\vr_2)\left[ \delta(\vr_1,\vr_2)-\g(\vr_1,\vr_2)\right]
\label{chikli}
\eea
where $A(\w)=4\D\ve/(\w^2-\D\ve^2)$. We will now use Eq. (\ref{chikli}), perform a similar approximation to $\L_s$ and simplify Eq. (\ref{lsseq}). In this way, as a first step, we obtain the so-called linearized HF approximation (LHF)\cite{lhfsg}
\bea
v_{\rm x}^{\rm LHF}(\vr_1)&=&\frac{\int d\vr_2\S_{\rm x}(\vr_1,\vr_2)\g(\vr_2,\vr_1)}{\g(\vr_1)}\nn\\
&&\!\!\!\!\!\!\!\!\!\!+\sum_{kk'}n_kn_{k'}\frac{\vf_k(\vr_1)\vf_{k'}(\vr_1)}{\g(\vr_1)}\bra \vf_k|D_\x|\vf_{k'}\ket,
\eea
where we have defined $D_\x(\vr,\vr')=v_\x(\vr)\d(\vr,\vr')-\S_\x(\vr,\vr')$ and $\g(\vr)=\g(\vr,\vr)$. The first term can be identified as the Slater potential $v^{S}_\x(\vr_1)$.\cite{slater} Making the further approximation of retaining only the diagonal elements of the second term we find the aforementioned KLI approximation\cite{kli92}
\bea
v_{\rm x}^{\rm KLI}(\vr_1)&=&v^{S}_\x(\vr_1)+\sum_kn_k\frac{|\vf_k(\vr_1)|^2}{\g(\vr_1)}\bra \vf_k|D_\x|\vf_k\ket.
\label{klipot}
\eea
The KLI potential, $v_{\rm x}^{\rm KLI}$, has shown to be a very good approximation to the true EXX potential. Performing a similar approximation to self-energies which contain correlation is less straight forward due to the extra energy dependence of a self-energy with correlation. However, using some additional approximation such equations have been derived by Casida.\cite{casida95} 

The KLI potential is undetermined up to the addition of a constant. However, applying Eq. (\ref{perdew2}) ensures that also $v_\x^{\rm KLI}\to 0$ as $r\to \infty$.\cite{kli92} We notice the similarity of the second term to the expression for the discontinuity within EXX (Eq. (\ref{perdew})). Indeed, it is easy to see that this is exactly the term that gives rise to steps in $v_\x^{\rm KLI}$.\cite{mkk11} These steps are of the same origin as the derivative discontinuity. The Slater potential misses any such features.  

\begin{figure}[t]
\includegraphics[width=8.2cm, clip=true]{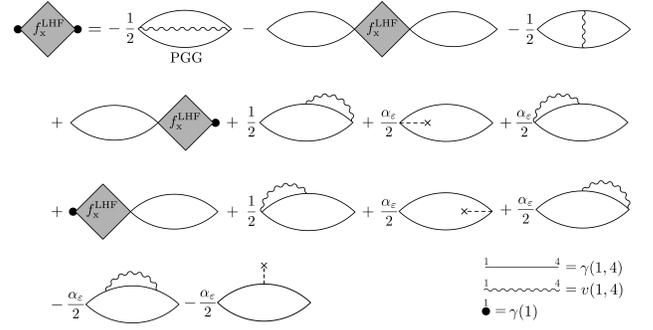}\\
\caption{Diagrammatic representation of the equation for the EXX kernel in the LHF approximation.}
\label{ediag}
\end{figure}

We will now study the TDEXX kernel in similar approximations (from now on called the 'EXX kernel'). The Slater or the KLI  approximations to the EXX kernel could be defined as the functional derivatives of the corresponding potentials. This would, however, lead to equations where the full $\chi_s$ has to be inverted. Instead, we start directly from the equation for the kernel. The left hand side of Eq. (\ref{fxceq}) contains two $\chi_s$ which both are replaced with Eq. (\ref{chikli}). The right hand side contains only energy differences between unoccupied and occupied KS states, thus permitting a similar CEDA. We find on the left hand side 
\bea
\int \!\!d\vr_2d\vr_3\,\chi_s(\vr_1,\vr_2;\w)f_{\rm x}(\vr_2,\vr_3;\w)\chi_s(\vr_3,\vr_4;\w)\qquad\qquad&&\nn\\
\!\!\!\!\!\!\!\!\approx A^2(\w)\g(\vr_1)f^{\rm LHF}_{\rm x}(\vr_1,\vr_4;\w)\g(\vr_4)\qquad\qquad\qquad\qquad\,\,\,\,\,\,\,&&\nonumber\\
\!\!\!\!\!\!\!\!-\,A^2(\w)\g(\vr_1)\int \!\!d\vr_3\,f^{\rm LHF}_{\rm x}(\vr_1,\vr_3;\w)\g^2(\vr_3,\vr_4)\quad\qquad\,\,\,\,\,\,&&\nonumber\\
\!\!\!\!\!\!\!\!-\,A^2(\w)\g(\vr_4)\int \!\!d\vr_2\,\g^2(\vr_1,\vr_2)f^{\rm LHF}_{\rm x}(\vr_2,\vr_4;\w)\quad\qquad\,\,\,\,\,\,&&\nonumber\\
\!\!\!\!\!\!\!\!+\,A^2(\w)\int \!\!d\vr_2d\vr_3\,\g^2(\vr_1,\vr_2)f^{\rm LHF}_{\rm x}(\vr_2,\vr_3;\w)\g^2(\vr_3,\vr_4).\,\,\,\,\,&&
\label{klilhs}
\eea
Details on the simplifications of the right hand side can be found in the Appendix. Here we simply state the result
\bea
R&=&-\frac{A^2(\w)}{2}\left[\g^2(\vr_4,\vr_1)v(\vr_1,\vr_4)\right.\nn\\
&&+\int \!\!d\vr_2\,\S_s^\x(\vr_1,\vr_2)\G(\vr_2,\vr_1;\vr_4)\nn\\
&&+\int \!\!d\vr_2\,\S_s^\x(\vr_4,\vr_2)\G(\vr_2,\vr_4;\vr_1)\nn\\
&&+\int \!\!d\vr_2d\vr_3\,\G(\vr_3,\vr_2;\vr_4)v(\vr_3,\vr_2)\G(\vr_3,\vr_2;\vr_1)\nn\\
&&-\a_\ve(\w)\left\{ B(\vr_1,\vr_4)\g(\vr_4,\vr_1) \right.\nn\\
&&\qquad\qquad\qquad+\left.\left.B(\vr_4,\vr_1)\g(\vr_4,\vr_1) \right\} \right].
\label{klirhs}
\eea
where
\bea
B(\vr_1,\vr_4)&=&\int  \!\!d\vr_2\,D_\x(\vr_1,\vr_2)\g(\vr_2,\vr_4)\nn\\
&&\!\!\!\!\!\!\!\!\!\!\!\!-\int  \!\!d\vr_2d\vr_3\,\g(\vr_1,\vr_2)D_\x(\vr_2,\vr_3)\g(\vr_3,\vr_4)
\eea
and $\G(\vr_3,\vr_2;\vr_1)=\g(\vr_2,\vr_1)\g(\vr_1,\vr_3)$ and $\a_\ve(\w)=(3\ve^2-\w^2)/2\ve^2$. We notice that only the density matrix with occupied orbitals appear. In Fig. \ref{ediag} Eqs. (\ref{klilhs}-\ref{klirhs}) is represented graphically which reveals better the structure of the equation. We see that the original equation is transformed from an equation where the Green's function is the basic variable to an equation where the density matrix is the basic variable. After division with the density we identify the first term to be the PGG kernel\cite{pgg96} 
\be
f_\x^{\rm PGG}(\vr_1,\vr_4)=-\frac{1}{2}\frac{\g^2(\vr_4,\vr_1)v(\vr_1,\vr_4)}{\g(\vr_1)\g(\vr_4)}
\ee
In analogy with the Slater potential the PGG kernel is obtained by ignoring the second term in the approximate response function of Eq. (\ref{chikli}). With the aim of studying the discontinuity of the kernel it is, however, essential to keep also the other terms as we will now see. We will make the further approximation of KLI, which means that we retain only diagonal terms in $k,k'$. Hence
\bea
f^{\rm KLI}(\vr_1,\vr_4;\w)&=&f_\x^{\rm PGG}(\vr_1,\vr_4)+f_\x^{\rm DD}(\vr_1,\vr_4;\w)
\label{klikern1}
\eea
where 
\bea
f^{\rm DD}_{\rm x}(\vr_1,\vr_4;\w)&=&\nn\\
&&\!\!\!\!\!\!\!\!\!\!\!\!\!\!\!\!\!\!\!\!\!\!\!\!\!\!\!\!\!\!\!\!\!\!\!\!-\sum_{k}^{\rm occ}\frac{|\vf_{k}(\vr_4)|^2}{\g(\vr_4)}\left[\frac{1}{2}\frac{\vf_{k}(\vr_1)}{\g(\vr_1)}\int \!\!d\vr_2\,\S_\x(\vr_1,\vr_2)\vf_{k}(\vr_2)\right.\nn\\
&&\quad\left.-\int d\vr_2f_{\rm x}^{\rm KLI}(\vr_1,\vr_2;\w)|\vf_k(\vr_2)|^2\right]\nn\\
&&\!\!\!\!\!\!\!\!\!\!\!\!\!\!\!\!\!\!\!\!\!\!\!\!\!\!\!\!\!\!\!\!\!\!\!\!-(\vr_1\leftrightarrow\vr_4)\nn\\
&&\!\!\!\!\!\!\!\!\!\!\!\!\!\!\!\!\!\!\!\!\!\!\!\!\!\!\!\!\!\!\!\!\!\!\!\!-\sum_{kp}^{\rm occ}\frac{|\vf_{k}(\vr_1)|^2}{\g(\vr_1)}\frac{|\vf_{p}(\vr_4)|^2}{\g(\vr_4)}\left[\frac{1}{2}\int\!\! d\vr_2d\vr_3\,\vf_{p}(\vr_2)\vf_{p}(\vr_3)\right.\nn\\
&&\qquad\,\,\,\,\,\times v(\vr_3,\vr_2)\vf_{k}(\vr_2)\vf_{k}(\vr_3)\nn\\
&&\!\!\!\!\!\!\!\!\!\!\!\!\!\!\!\!\!\!\!\!\!\!\!\!\!\left.+\int\!\!d\vr_2d\vr_3\,|\vf_{k}(\vr_2)|^2f_{\rm x}^{\rm KLI}(\vr_2,\vr_3;\w)|\vf_{p}(\vr_3)|^2\right]\nn\\
&&\!\!\!\!\!\!\!\!\!\!\!\!\!\!\!\!\!\!\!\!\!\!\!\!\!\!\!\!\!\!\!\!\!\!\!\!-\a_\ve(\w)\sum_k^{\rm occ}\frac{|\vf_k(\vr_1)|^2}{\g(\vr_1)}\frac{|\vf_k(\vr_4)|^2}{\g(\vr_4)}\bra\vf_k|D_\x|\vf_k\ket\nonumber\\
&&\!\!\!\!\!\!\!\!\!\!\!\!\!\!\!\!\!\!\!\!\!\!\!\!\!\!\!\!\!\!\!\!\!\!\!\!+\frac{\a_\ve(\w)}{2}\sum_k^{\rm occ}\frac{|\vf_k(\vr_4)|^2}{\g(\vr_4)}\left[ v_\x(\vr_1)\frac{|\vf_k(\vr_1)|^2}{\g(\vr_1)}\right.\nn\\
&&\left.-\frac{\vf_k(\vr_1)}{\g(\vr_1)}\int d\vr_2\S_{\x}(\vr_1,\vr_2)\vf_k(\vr_2)\right]\nn\\
&&\!\!\!\!\!\!\!\!\!\!\!\!\!\!\!\!\!\!\!\!\!\!\!\!\!\!\!\!\!\!\!\!\!\!\!\!+(\vr_1\leftrightarrow\vr_4)
\label{klikern2}
\eea
We note that the frequency dependence is not completely eliminated by the KLI approximation. Terms which behave as $\w^2$ remains. The KLI kernel is undetermined up to the addition of two functions $g(\vr,\w)$ and $g(\vr',\w)$, as in the case of the exact EXX kernel. In the previous section we derived conditions that should be imposed on Eq. (\ref{fxceq}) in order to fix these function. By performing a KLI approximation to the same equation we find 
\bea
0&=&\frac{1}{2}\frac{|\vf_{\rm H}(\vr_1)|^2}{\g(\vr_1)}v_\x(\vr_1)-\frac{\vf_{\rm H}(\vr_1)}{\g(\vr_1)}\int \!\!d\vr_2\,\S_\x(\vr_1,\vr_2) \vf_{\rm H}(\vr_2)\nn\\
&&\!\!\!\!\!\!\!+\int \!\!d\vr_2\,|\vf_{\rm H}(\vr_2)|^2f_\x^{\rm KLI}(\vr_2,\vr_1)\nn\\
&&\!\!\!\!\!\!\!-\sum_k^{\rm occ}\frac{|\vf_k(\vr_1)|^2}{\g(\vr_1)}\left[\int \!\!d\vr_2d\vr_3\,|\vf_{\rm H}(\vr_2)|^2f_\x^{\rm KLI}(\vr_2,\vr_3)|\vf_k(\vr_3)|^2\right.\nn\\
&&\!\!\!\!\!\!\!\left.+\frac{1}{2}\int \!\!d\vr_2d\vr_3\,\vf_k(\vr_2)\vf_k(\vr_3)v(\vr_2,\vr_3) \vf_{\rm H}(\vr_2)\vf_{\rm H}(\vr_3)\right]
\eea
This condition together with 
\bea
0&=&\int \!\!d\vr_2d\vr_3\,|\vf_{\rm H}(\vr_2)|^2f_\x^{\rm KLI}(\vr_2,\vr_3)|\vf_{\rm H}(\vr_3)|^2\nn\\
&&+\frac{1}{2}\int \!\!d\vr_2d\vr_3\,|\vf_{\rm H}(\vr_2)|^2v(\vr_2,\vr_3)|\vf_{\rm H}(\vr_3)|^2
\eea
arising from Eq. (\ref{muxcdeltandeltan}) allows one to remove all the terms containing the HOMO orbital in $f_\x^{\rm DD}$. At the same time $f^{\rm KLI}_\x$ is uniquely determined. Notice that we have not considered the frequency dependence here.

The discontinuity in the KLI approximation can also be determined
\bea
g_\x(\vr_1)&=&\nn\\
&&\!\!\!\!\!\!\!\!\!\!\!\!\!\!\!\!\!\!\!\!\!\frac{1}{2}\frac{|\vf_{\rm L}(\vr_1)|^2}{\g(\vr_1)}v_\x(\vr_1)-\frac{\vf_{\rm L}(\vr_1)}{\g(\vr_1)}\int \!\!d\vr_2\,\S_\x(\vr_1,\vr_2) \vf_{\rm L}(\vr_2)\nn\\
&&\!\!\!\!\!\!\!\!\!\!\!\!\!\!\!\!\!\!\!\!\!+\int \!\!d\vr_2\,|\vf_{\rm L}(\vr_2)|^2f^-_\x(\vr_2,\vr_1)\nn\\
&&\!\!\!\!\!\!\!\!\!\!\!\!\!\!\!\!\!\!\!\!\!-\sum_k\frac{|\vf_k(\vr_1)|^2}{\g(\vr_1)}\left[\int \!\!d\vr_2d\vr_3\,|\vf_{\rm L}(\vr_2)|^2f^-_\x(\vr_2,\vr_3)|\vf_k(\vr_3)|^2\right.\nn\\
&&\!\!\!\!\!\!\!\!\!\!\!\!\!\!\!\!\!\!\!\!\!-\int \!\!d\vr_3\,g_\x(\vr_3)|\vf_k(\vr_3)|^2\nn\\
&&\!\!\!\!\!\!\!\!\!\!\!\!\!\!\!\!\!\!\!\!\!\left.+\frac{1}{2}\int \!\!d\vr_2d\vr_3\,\vf_k(\vr_2)\vf_k(\vr_3)v(\vr_2,\vr_3) \vf_{\rm L}(\vr_2)\vf_{\rm L}(\vr_3)\right].
\eea
This equation shows that the sum of the additional terms beyond the PGG can be seen as the explicit inclusion of the discontinuity.

We will now briefly discuss how the KLI equations are solved in practice. As suggested in Ref. \onlinecite{kli92} the KLI potential can be determined by first multiplying Eq. (\ref{klipot}) with $|\vf_k(\vr_1)|^2$, where $k$ is occupied, and then integrate with respect to $\vr_1$. This leads to a matrix equation determining the constants $\int |\vf_k|^2 v_\x^{\rm KLI}$, appearing on the right hand side of Eq. (\ref{klipot}). When these constants are known $v_\x^{\rm KLI}(\vr_1)$ is easily calculated. The same idea is here used for calculating the KLI kernel (Eq. (\ref{klikern1}-\ref{klikern2})). Hence, we first multiply with $|\vf_k(\vr_1)|^2$ and $|\vf_p(\vr_4)|^2$ and then integrate with respect to $\vr_1$ and $\vr_4$. This allows us to calculate the constants $\int |\vf_k|^2f^{\rm KLI}_\x|\vf_p|^2$ via a matrix equation. With these constants known it is easy to derive another equation for the vectors $\int d\vr_2f^{\rm KLI}_\x(\vr_1,\vr_2;\w)|\vf_p(\vr_2)|^2$. Knowing these constants and vectors we can finally calculate $f^{\rm KLI}_\x(\vr_1,\vr_2;\w)$.  
\section{Numerical results and analysis}
In this section we present numerical results for one-dimensional (1D) soft-Coulomb systems. Such model systems are often
used to study qualitative features similar to those arising in 3D molecules. In this model the Coulomb interaction is 'softened' such that the electron-electron interaction is given by $1/\sqrt{(x_1-x_2)^2+1}$, where $x_1$ and $x_2$ are the coordinates of electron 1 and 2, respectively. Also the nuclear potentials are softened and given by $1/\sqrt{(x_1-Q)^2+1}$, where $Q$ is the position of the nucleus.

All calculations have been performed using cubic splines as basis functions, uniformly distributed along the $x$-axis. Details on the numerical approach can be found in Ref. \onlinecite{hvb07}.
\subsection{Diatomic molecules at dissociation}
A molecule that dissociates is one of the hardest systems to describe within DFT. This is especially true when the molecule consists of open-shell atoms and has covalent bonds. In this case the system is strongly correlated and cannot be described with an exchange-only theory. Applying the EXX approximation leads to largely underestimated total energies and fractionally charged dissociation fragments. The latter, so-called delocalization problem is due to a missing derivative discontinuity at {\em odd} integer particle numbers in the EXX functional. 

Molecules composed of closed-shell atoms can also be difficult to describe at dissociation using standard functionals giving rise to similar fractional charge problems. In the case of closed-shell fragments the functional needs to have a derivative discontinuity at {\em even} integers. As we have seen in the previous section such a discontinuity is contained already at the EXX level. This allows us to study the fractional charge/discontinuity problem within the EXX functional. 

In order to see when the discontinuity is needed for describing a stretched molecule one can study its constituent atoms independently. Since the unoccupied KS levels have no physical significance it may happen that the KS affinity of one of the atoms is larger than the ionization energy of the other atom. Enforcing the aufbau principle a step over the atom with the large affinity is needed when the atoms are described as a combined system at infinite separation. An example of such a system is the model HeBe$^{2+}$ molecule, in which the nuclear strength of the He nucleus is set to 2.5. This gives an ionization energy $I=1.147$ and a KS affinity $A_s=0.494$. The Be$^{2+}$ nucleus is set to 4.5 giving $I=2.836$ and $A_s=1.673$. We now see that $A_s$ of the Be$^{2+}$ is larger than $I$ of the He atom and a step of minimum 0.526 is required at infinite separation. 

\begin{figure}[t]
\includegraphics[width=8.5cm, clip=true]{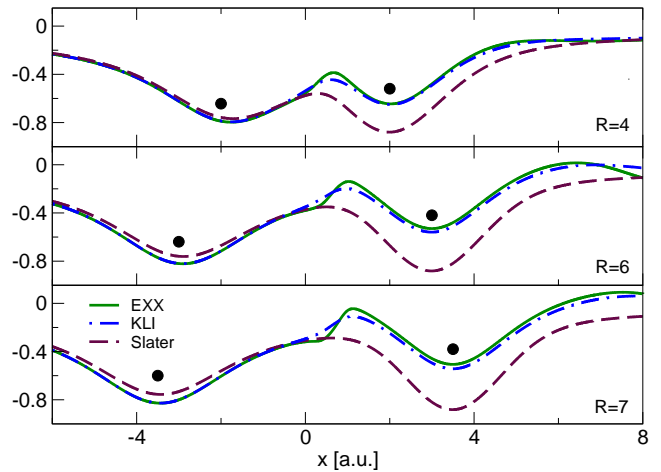}
\caption{The black circles show the position of two atoms at different separations $R$. The left with nuclear strength 2.5 and the right 4.5. In total there are 4 electrons. Due to the missing step in the PGG approximation the density dissociates with fractional charges on the atoms. In this case the effect is rather small, the error being around 10\%. }
\label{potent}
\end{figure}

Figure \ref{potent} shows the EXX, the KLI and the Slater potentials of the HeBe$^{2+}$ system at different separation $R$. Both the EXX and the KLI have a step as predicted, being somewhat sharper in the EXX. The step insures integer charges on the atoms, something which we have also verified numerically. The Slater potential which lacks the step gives an excess of 0.1 charges on the Be$^{2+}$ atom. At large separation, the difference between the Slater and the KLI/EXX potential is almost exactly a constant over the Be$^{2+}$ atom showing the close relation between the second term of Eq. (\ref{klipot}) and the discontinuity.

\begin{figure}[t]
\includegraphics[width=8.5cm, clip=true]{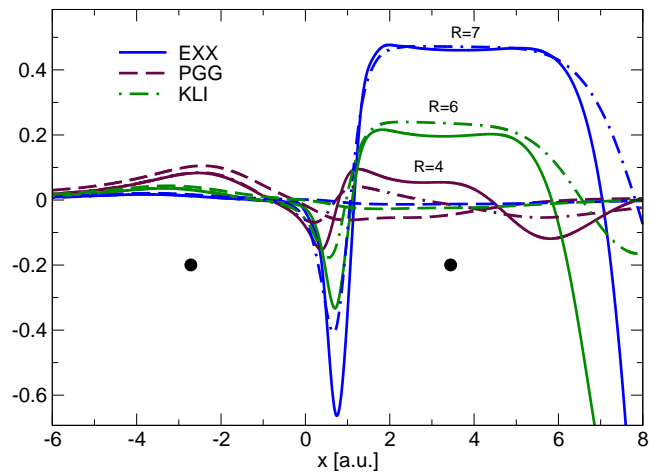}
\caption{The same systems as in Fig. \ref{potent}. The quantity $\int d\vr' \F_q(\vr')f_{\x}(\vr',\vr,\w=0)$ is plotted where $\F_q$ is the excitation function of the HOMO-LUMO excitation being a charge transfer excitation. }
\label{potent11}
\end{figure}

In Fig. \ref{potent11} the kernel of the same system is shown. The quantity that is plotted is 
\be
\int d\vr' \F_q(\vr')f_{\x}(\vr',\vr,\w=0),
\label{fqfxc}
\ee 
which is equal to the induced change in $v_\x$ given that the density change is equal to $\F_q$. Here $\F_q$ is set to the excitation function corresponding to the HOMO-LUMO transition thus representing a charge-transfer excitation. We see that even though $\F_q$ is vanishingly small due to the exponentially decreasing overlap of the HOMO and at the LUMO orbitals the kernel shows large structures in terms of peaks and a plateau that grows with separation. This structure which is missing in the PGG kernel comes entirely from $f^{DD}_\x$ which contains the effects of the discontinuity. In the next subsection we will see that the discontinuity has very similar shape. Integrating Eq. (\ref{fqfxc}) with another $\F_q$ gives $\bra \F_q|f_\x|\F_q\ket$ which is proportional to the correction to the charge-transfer excitation energy in the adiabatic single-pole approximation. While in the PGG this correction quickly tends to zero it grows with separation in both EXX and KLI. On the other hand, the matrix element of $f_\x$ calculated with an $\F_q$ of an excitation localized on the Be atom converges to the value of the isolated atom, unaffected by the discontinuity. Thus, {\em locally} the discontinuity merely adds an irrelevant function $g_\xc$ to the kernel. We will discuss excitation energies more in Sec. IIID. 



\subsection{Ensemble $v_\x$ and $f_\x$}
As discussed previously the property a functional needs to have in order produce a step in $v_\xc$ in a stretched molecule is 
the derivative discontinuity. The discontinuity of a given functional can be studied if the domain of densities is extended to ensemble densities of the kind discussed in Sec. IIA-B. In the following we will calculate the ensemble $v_\x$ and $f_\x$ and their discontinuities and relate them to the structures seen in the stretched molecules. 
\begin{figure}[t]
\includegraphics[width=8.5cm, clip=true]{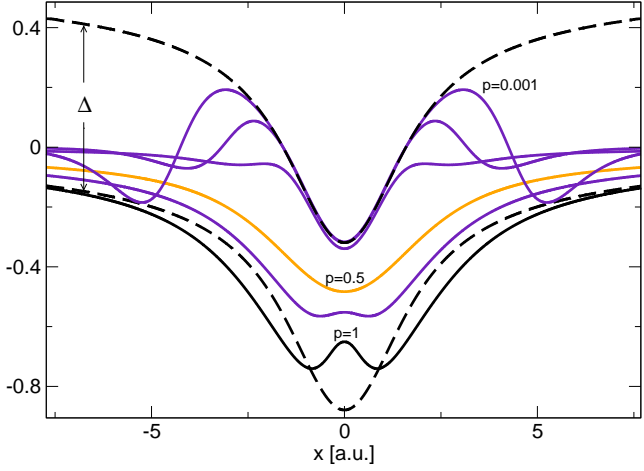}
\caption{The ensemble EXX potential of Be$^{2+}$ for different values of $p$ such that $4 \leqslant N \leqslant 2$. The dashed curves are obtained by calculating the 2 electron potential from the two different limits $N\to 2^\pm$.}
\label{potent1}
\end{figure}

\begin{figure}[t]
\includegraphics[width=8.5cm, clip=true]{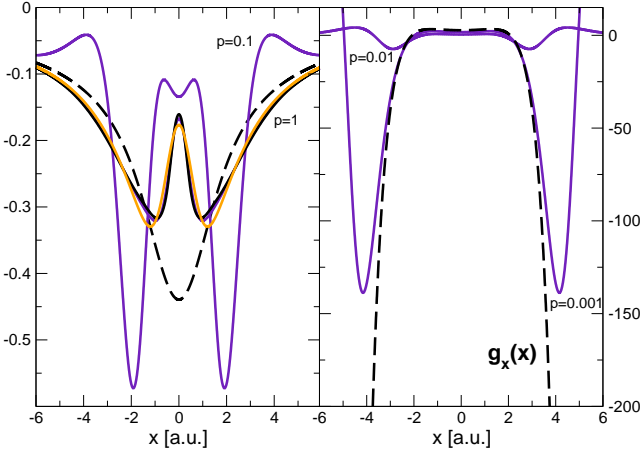}
\caption{The ensemble EXX kernel of Be$^{2+}$ for different values of $p$ such that $4 \leqslant N \leqslant 2$. Left panel shows $p=1,0.8,0.5,0.1$ and the left panel shows $p=0.01,0.001$. The dashed curves in the different panels are obtained by calculating the 2 electron potential from the two different limits $N\to 2^\pm$.}
\label{potent2}
\end{figure}

Eqs. (\ref{lsseq})-(\ref{fxceq}) was generalized to ensembles as discussed in Sec. III and solved numerically together with the conditions of Eq. (\ref{perdew2}) and Eq. (\ref{disckleintd}) (before the limit $N\to N_0^+$ is taken). The discontinuities were also calculated (Eqs. (\ref{perdew}) and (44)). Figure \ref{potent1} shows the EXX potential of Be$^{2+}$ for different values of $p$. By varying $p$ between 0 and 1 the number of particles change between 2 and 4. The dashed curves represent $v_\x$ for 2 electrons but calculated from the two different limits $N\to 2^+$ and $N\to 2^-$. The difference between the curves is the discontinuity $\D_\x$, exactly given by Eq. (\ref{perdew}). This was also verified numerically. For $p<1$ we find a step in $v_\x$ that becomes sharper and sharper and moves further away from the center as $p\to 0^+$. This structure is similar to the structure over the Be$^{2+}$ atom in the molecule for EXX and KLI, thus confirming the close relation between the discontinuity of the ensemble potential and the step during dissociation.
\begin{figure}[b]
\includegraphics[width=8.5cm, clip=true]{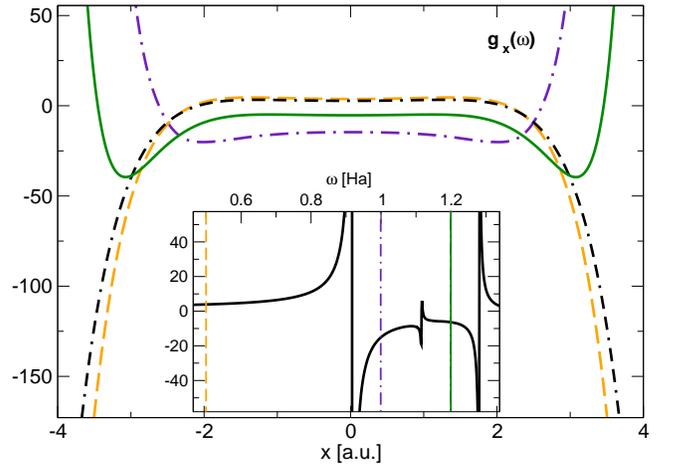}
\caption{The discontinuity of the EXX kernel calculated at different frequencies. The inset shows $\int |\vf|^2 f_\x (\w)|\vf|^2$ and at which frequency the discontinuity is evaluated.}
\label{potent3}
\end{figure}

In Fig. \ref{potent3} we have plotted the corresponding static ensemble kernel 
\be
F_{\rm H}(\vr)=\int d\vr' f_\x(\vr,\vr')|\vf_{\rm H}(\vr')|^2
\ee
and we notice that
\bea
F^+_{\rm H}(\vr)&=&\int d\vr' f^-_\x(\vr,\vr')|\vf_{\rm H}(\vr')|^2\nn\\
&&+\int d\vr' g_\x(\vr')|\vf_{\rm H}(\vr')|^2+g_\x(\vr)
\label{plotfx}
\eea
The dashed lines correspond to the kernel calculated from the two different limits $N\to 2^+$ and $N\to 2^-$. Here we see a rather large difference which amounts to the discontinuity $g_\x$ and the constant $\int  g_\x|\vf_{\rm H}|^2$ according to Eq. (\ref{plotfx}). The diverging behavior that was deduced from the analysis in Sec. IIB is thus confirmed for the EXX kernel. Before the limit is reached the ensemble kernel displays sharp peaks and a plateau-like structure very similar to the ones observed in the stretched molecule of the EXX and KLI. The fact that the discontinuity diverges is exactly what allows the kernel of the combined system to capture long range charge-transfer excitations. Figure \ref{potent3} shows the discontinuity at $\w=0$, $\w=0.5$ (below the first excitation energy), $\w=1.0$ and at $\w=1.2$. The discontinuity obtained from Eq. (\ref{disckleintd}) contains poles at the energy differences between the LUMO and rest of the unoccupied and occupied states (see inset of Fig. \ref{potent3}). Only the LUMO to the occupied states correspond to KS excitation energies contained also in $\chi_s$. 
We see that the frequency changes the strength of the divergency of the discontinuity. This is expected since the same kernel has to describe all charge transfer states which all have different excitation functions decaying at different rate. 
\subsection{Field-counteracting effect}
In Refs. \onlinecite{gsgbsck99,kkp04} it was found that when applying a weak electric field to a chain of hydrogen molecules the EXX potential exhibits a step structure, counter-acting the applied field. It was then shown that calculating the polarizability by taking the difference between densities with and without field the values where strongly dependent of this property. The field-counteracting effect is an exchange effect when the constituent molecules are closed shell as in the case of an H$_2$-chain. This allows us here to study this effect and give it a perspective in terms of the discontinuity of the kernel. 
\begin{figure}[t]
\includegraphics[width=8.5cm, clip=true]{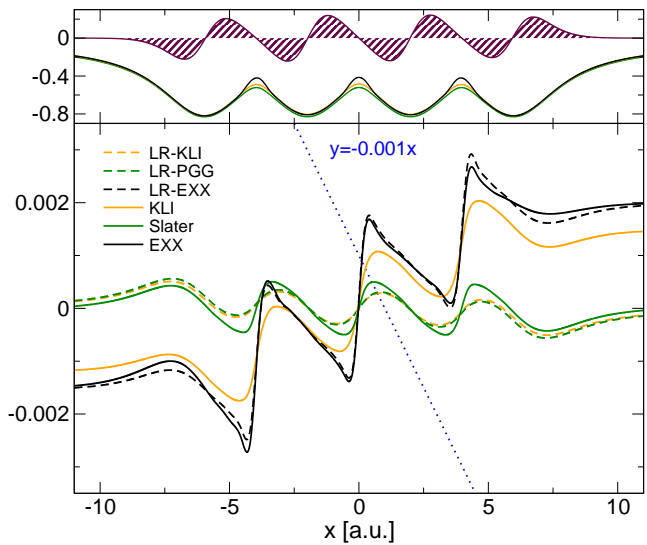}\\
\caption{Upper panel: The EXX potential (EXX, PGG and KLI) of a chain of four H$_2$ molecules. The density difference obtained when applying a linear potential is shown as the shaded top curve. Lower panel: The EXX potential differences when applying a linear potential as calculated from the kernel (dashed lines) and as the difference between two independent self-consistent KS calculations (solid lines). Blue dotted curve is the applied linear potential. }
\label{ediag4}
\end{figure}
\begin{figure}[t]
\includegraphics[width=8.5cm, clip=true]{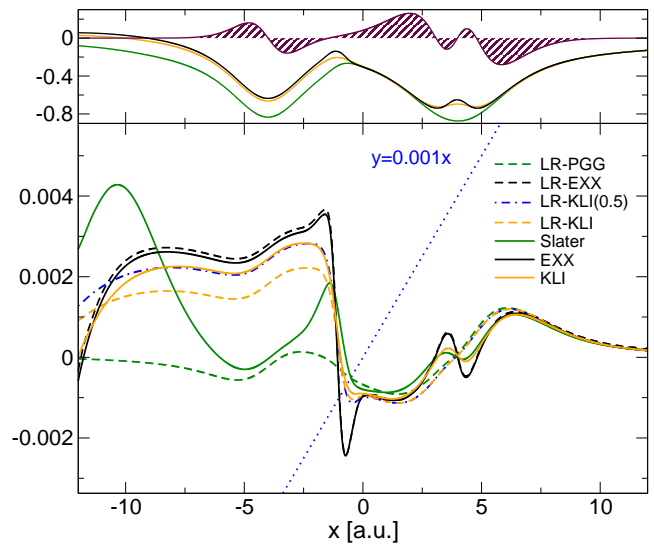}\\
\caption{Upper panel: The EXX potential (EXX, PGG and KLI) of a He-Be$^2+$ dimer. The density difference obtained when applying a linear potential is shown as the shaded top curve. Lower panel: The EXX potential differences when applying a linear potential as calculated from the kernel (dashed lines) and as the difference between two independent self-consistent KS calculations (solid lines). The dashed-dot curve is obtained by setting $\a=0.5$. Blue dotted curve is the applied linear potential. }
\label{ediag4b}
\end{figure}

We have studied a chain of four H$_2$-molecules. Fig. \ref{ediag4} (upper panel) shows the PGG/KLI/EXX potentials of this chain as well as the density difference obtained by applying a weak constant electric field with potential $V=0.001x$. In the lower panel the EXX potential differences ($\D v_\x$) are displayed with full lines. We clearly see a step structure counteracting the applied field in the EXX and the KLI. The KLI somewhat underestimates the incline and the PGG misses it completely, which is related to the missing discontinuity in the Slater potential. 

An alternative way to calculate the potential differences is to use 
\be
\d v_{\xc}(\vr) = \int d\vr' f_{\xc}(\vr,\vr')\d n(\vr'),
\ee 
exact when the applied field is infinitesimally small. The results can be found as the dashed curves in Fig. \ref{ediag4}. Note here that only the full EXX is expected to give an identical result to the potential difference calculations since only in that case does the kernel represent the true functional derivative. The KLI gives a result very similar to PGG. This has to do with the fact that the system is symmetric and was pointed out already in Ref. \onlinecite{gb01}. 

Figure \ref{ediag4} shows the same curves as in the case of the H$_2$ chain but for the asymmetric HeBe$^{2+}$ system. Again a step is found in EXX and KLI but missing in PGG. We also see that it is possible to tune the KLI result by changing the frequency such that the response calculation coincides with the potential differences. The frequency dependence of the KLI can 
thus change the strength of the discontinuity. It is clear that the step structures found in $v_\x$ can be seen as a consequence of the discontinuity in $f_\x$. 
\subsection{Excitation energies}
The LR-TDDFT equation is often compared to the Bethe-Salpeter equation (BSE). The BSE contains a four-point kernel and an interacting Green's function as input whereas the LR-TDDFT equation requires only a two-point kernel and a non-interacting KS Green's function as input. For molecular valence excitations a KS Green's function can be a better starting point than an interacting Green's function. The latter contains energy differences with respect to addition and removal energies that needs to be corrected with the particle-hole interaction, and part of this is already contained in the KS excitation energies due to the $1/R$ behavior of the KS potential. There are, however, excitations for which a full G serves as a better starting point. Among those excitations belong long-range charge-transfer and inner-shell excitations. 
It is therefore not surprising that a theory based on replacing the interacting $G$ with a non-interacting $G_s$ has difficulties in accurately capturing these excitations.\cite{hvb09,hg12} We will see examples of this problem shortly but before that we will investigate the single-pole approximation (SPA) within LR-TDDFT. The SPA ignores off-diagonal elements in the Casida matrix and expands the diagonal elements around the KS excitation energies. One obtains
\be
\Omega_q=\w_{q}+2\bra \F_q|v+f_\x(\w_q)|\F_q \ket.
\ee
One can also show that this approximation yields excitation energies identical to the ones obtained from first-order G\"orling-Levy perturbation theory.\cite{gs99,gl93} Within the SPA we can thus calculate excitation energies not subjected to errors that can arise when carrying out improper partial summations.

\begin{figure}[t]
\includegraphics[width=8.5cm, clip=true]{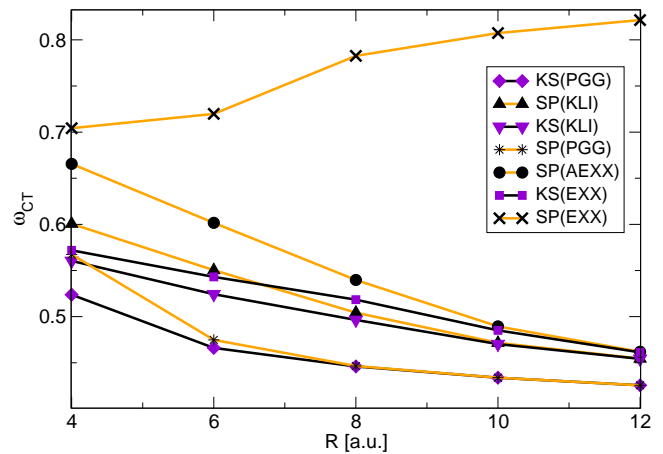}
\caption{The first charge-transfer excitation energy for the system in Fig. \ref{potent5} evaluated in the SPA. Only the fully frequency dependent kernel is able to give a correction to the KS eigenvalue difference at large $R$.}
\label{potent4}
\end{figure}

\begin{figure}[t]
\includegraphics[width=8.5cm, clip=true]{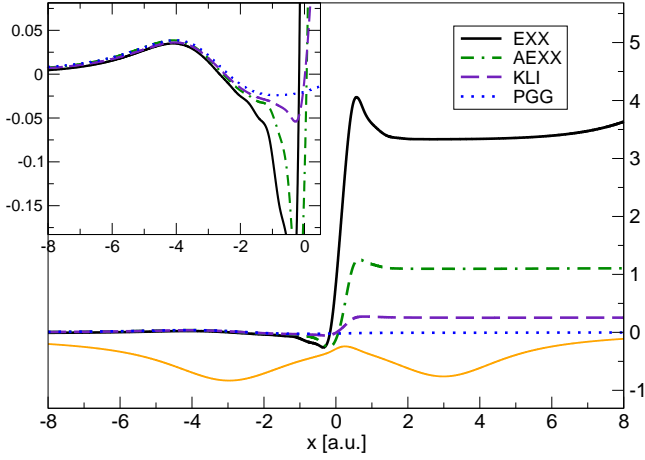}
\caption{The kernel of a stretched HeBe$^{2+}$ molecule for which the charge on the Be$^{2+}$ atom is set to 3 a.u. In this case the step in $v_\x$ vanishes in the dissociation limit but the step in $f_\x$ grows and eventually diverge.}
\label{potent5}
\end{figure}

Figure \ref{potent4} shows the first charge-transfer excitation energy of the HeBe$^{2+}$ system. We have decreased the charge on the Be$^{2+}$ atom to 3 a.u, in order to remove the step in $v_\x$. In this way the KS excitation energies are not artificially shifted and the effect of the discontinuity comes entirely from the kernel (see Fig. \ref{potent5}). In the SPA we see that only the fully frequency dependent kernel is able give a finite correction to the KS excitation energy in the dissociation limit. Figure \ref{potent5} indeed shows that the step is strongly enhanced at finite frequency. The adiabatic approximation works rather well for small $R$ but then deviates more and more and finally joins the KS eigenvalue difference. The KLI kernel behaves similarly to the AEXX and the PGG is seen to even more rapidly converge to the KS energy. 

\begin{figure}[t]
\includegraphics[width=8.5cm, clip=true]{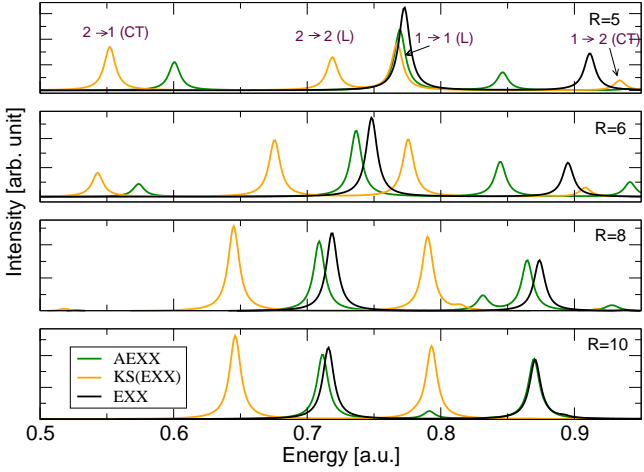}
\caption{The spectrum of HeBe$^{2+}$ at different separation $R$. The KS charge-transfer (CT) and the local (L) excitations are identified at $R=5$.}
\label{potent6}
\end{figure}

If we solve the full LR-TDDFT equation we find the results of Fig. \ref{potent6}. The first two charge-transfer (CT) excitations as well as one local (L) excitation on each atom are identified. With the AEXX we find the same number of peaks but shifted to higher energies. The charge transfer excitations are also here seen to reduce to the KS eigenvalue 
differences in the dissociation limit, even slightly faster than in the SPA. The local excitations are converging to the values found for the isolated atoms. Since there are only two electrons on each atom the AEXX and the full EXX are identical in the dissociation limit with respect to local excitations. This is indeed also what we see at $R=10$, which confirms the fact that the large step in $f_\x$ (Fig. \ref{potent5}) does not influence the local excitations. The effect of the frequency dependence on the CT excitations appears to be to remove them completely from the spectrum. A smeared structure can be seen for $R=5$ at $\w=0.6$ but if we reduce the $\d$-width this structure becomes even more broadened, which suggest that this is not a proper excitation. A similar disappearance of peaks happen in the case of inner-shell excitations as we will now discuss.      

To investigate inner-shell transitions we study a 4 electron 1D atom with the nuclear charge set to 4.5 a.u. The KS eigenvalue difference for the first inner-shell transition is around 1.3 a.u. In the SPA the different approximations give a correction of 0.1624 (EXX), 0.0132 (AEXX) 0.0152 (KLI), -0.001755 (PGG). PGG is the only approximation that gives a negative contribution, which also is very small. KLI and AEXX gives a correction an order of magnitude larger than PGG  and the frequency dependent EXX yet another order of magnitude. Clearly a frequency dependence is important in this case and this is true for all inner-shell excitation energies. The fact that PGG differs so much from KLI also suggests that the discontinuity is of importance for these excitations. 

\begin{figure}[t]
\includegraphics[width=8.5cm, clip=true]{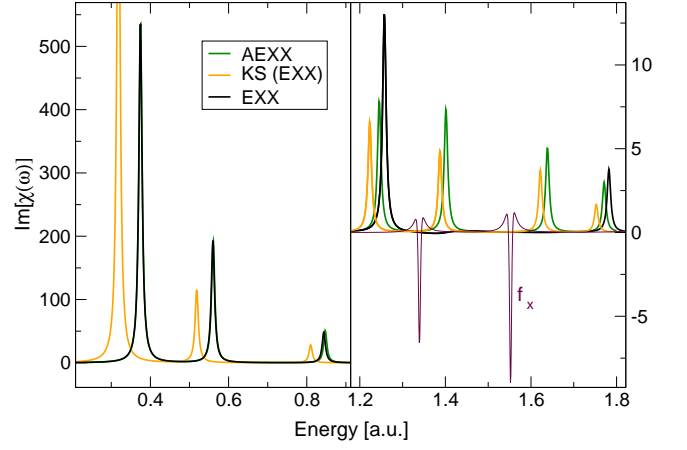}
\caption{Spectrum of a four electron 1D Be atom. At low energy the EXX and AEXX are very similar but start to differ at higher energies. In the EXX all inner-shell excitations are missing due to the double pole structure that can be found in the kernel close to every inner-shell excitation.}
\label{potent7}
\end{figure}

If we instead solve the full LR-TDDFT equations with the EXX kernel we find that all inner-shell excitations are missing from the spectra (Fig. \ref{potent7}). This fact was already discovered in Ref. \onlinecite{hvb09} and here we see that also 1D systems exhibit this behavior. The reason for this is the double-pole structure of $f_\x$ which in turn is related to the double inversion of the KS response function, which has zero eigenvalues at these frequencies. Double-poles can lead to a non-causal response function with poles symmetrically located in upper/lower half of the complex plane as shown in Ref. \onlinecite{hvb09}. The exact kernel must have a strong frequency dependence at these energies but  further analysis shows that a single-pole structure is the correct one. 

Figure \ref{potent7} also shows the result obtained using the AEXX, showing that at lower energies EXX and AEXX are very close for this system.

\section{Conclusions}
In this paper we have analyzed the integer discontinuity of the XC kernel by extending its domain of 
densities to ensemble densities that integrate to non-integer particle numbers. It was found
that in order to completely determine the discontinuity the ensemble densities had to
be general enough to allow for changes of particle numbers in time. 

The discontinuity of the ensemble kernel was evaluated showing a diverging spacial dependence and a pole 
structure in the frequency. Such strong features have, however, no affect on the calculated 
spectra at integer particle numbers. On the other hand, in a combined system, such as a stretched molecule, 
where the discontinuity shows up on only a part of the system the discontinuity can have a large effect. This implies that locally such behavior does not affect the excitations but for excitations involving transfer of charge the affect can be large.

A numerical study was performed within the EXX approximation. Two different approximations to 
the EXX kernel was derived. The PGG approximation which has been used previously and a new KLI-type of 
approximation solved numerically for the first time here. The KLI approximation captures explicitly the 
discontinuity of the full adiabatic EXX approximation. It was also shown that an additional term in the KLI 
approximation could be used to tune the strength of the discontinuity. 

By comparing the PGG, KLI, AEXX and EXX we were able to identify properties that are dependent on the 
discontinuity of the functional. For example the field counteracting effect in a molecule chain can be seen as a consequence of the discontinuity of the kernel. For charge transfer excitations the discontinuity provides the 
diverging behavior needed to compensate for the vanishing overlap of the KS orbitals. Also local inner-shell 
excitations are strongly modified when incorporating the discontinuity. It was also shown that none of the approximation studied here has the proper frequency dependence to fully account for these excitations. There is thus a need to go beyond exact-exchange for accurately calculating spectral properties within TDDFT.
\appendix
\section{Expanded derivations}
In the EXX, the right hand side of Eq. (\ref{fxceq}) can be separated into a vertex term (third diagram of Fig. \ref{ediag1}) and self-energy terms. For the vertex we find the following two terms:
\bea
V_1&=&\frac{4}{\w^2-\varepsilon^2}\left[\g^2(\vr_4,\vr_1)v(\vr_4,\vr_1)\right.\qquad\nn\\
&&+\int d\vr_2d\vr_3\g(\vr_4,\vr_2)\g(\vr_4,\vr_3)v(\vr_3,\vr_2)\nn\\
&&\qquad\times\g(\vr_2,\vr_1)\g(\vr_3,\vr_1)\nn\\
&&+\int d\vr_2\S(\vr_1,\vr_2)\g(\vr_2,\vr_4)\g(\vr_4,\vr_1)\nn\\
&&\left.+\int d\vr_2\S(\vr_4,\vr_2)\g(\vr_2,\vr_1)\g(\vr_4,\vr_1)\right]
\eea
and
\bea
V_2&=&-4\frac{\w^2+\varepsilon^2}{(\w^2-\varepsilon^2)(\w^2-\varepsilon^2)}\left[\int d\vr_2d\vr_3\g(\vr_4,\vr_2)
\right.\nn\\
&&\qquad\qquad\times\g(\vr_4,\vr_3)v(\vr_3,\vr_2)\g(\vr_2,\vr_1)\g(\vr_3,\vr_1)\nn\\
&&+\int d\vr_2\S(\vr_1,\vr_2)\g(\vr_2,\vr_4)\g(\vr_4,\vr_1)\nn\\
&&+\int d\vr_2\S(\vr_4,\vr_2)\g(\vr_2,\vr_1)\g(\vr_4,\vr_1)\nn\\
&&\left. -\int d\vr_2\S(\vr_1,\vr_2)\g(\vr_2,\vr_4)\d(\vr_4,\vr_1)\right]
\eea
The self energy terms can be summarized as
\bea
T_1&=&4\frac{\w^2+\varepsilon^2}{(\w^2-\varepsilon^2)(\w^2-\varepsilon^2)}\left[\int d\vr_2\vr_3\g(\vr_1,\vr_2)\right.\nn\\
&&\quad\left.\times\D(\vr_2,\vr_3)\g(\vr_3,\vr_4)\left[\delta(\vr_1,\vr_4)-\g(\vr_4,\vr_1)\right]\right]
\eea
\bea
T_2=T_3&=&-4\frac{\g(\vr_4,\vr_1)}{\w^2-\varepsilon^2}\left[\int \!d\vr_2\D(\vr_1,\vr_2)\g(\vr_2,\vr_4)\right.\nn\\
&&-\left.\int\! d\vr_2d\vr_3\g(\vr_1,\vr_2)\D(\vr_2,\vr_3)\g(\vr_3,\vr_4)\right]
\eea
\bea
T_4=T_5&=&-4\frac{\g(\vr_4,\vr_1)}{\w^2-\varepsilon^2}\left[\int \!d\vr_2\D(\vr_4,\vr_2)\g(\vr_2,\vr_1)\right.\nn\\
&&-\left.\int\! d\vr_2d\vr_3\g(\vr_1,\vr_2)\D(\vr_2,\vr_3)\g(\vr_3,\vr_4)\right]
\eea
\bea
T_6&=&-\frac{\w^2+\varepsilon^2}{(\w^2-\varepsilon^2)(\w^2-\varepsilon^2)}\g(\vr_4,\vr_1)\left[\D(\vr_1,\vr_4)\right.\nn\\
&&+\int d\vr_2d\vr_3\g(\vr_1,\vr_2)\D(\vr_2,\vr_3)\g(\vr_3,\vr_4)\nn\\
&&-\int d\vr_2\g(\vr_1,\vr_2)\D(\vr_2,\vr_4)\nn\\
&&\left.-\int d\vr_2\D(\vr_1,\vr_2)\g(\vr_2,\vr_4)\right]
\eea
Summing up everything we find Eq. (\ref{klirhs}).
\newpage
\bibliography{../ref}

\end{document}